\documentclass{sigchi}

\CopyrightYear{2024}
\setcopyright{acmlicensed}
\doi{https://doi.org/10.1145/3313831.XXXXXXX}
\isbn{978-1-4503-6708-0/20/04}
\conferenceinfo{CHI'20,}{April  25--30, 2020, Honolulu, HI, USA}
\acmPrice{\$15.00}

\toappear{
Permission to make digital or hard copies of all or part of this work
for personal or classroom use is granted without fee provided that
copies are not made or distributed for profit or commercial advantage
and that copies bear this notice and the full citation on the first
page. Copyrights for components of this work owned by others than Ghamut Corporation
must be honored. Abstracting with credit is permitted. To copy
otherwise, or republish, to post on servers or to redistribute to
lists, requires prior specific permission and/or a fee. Request
permissions from \href{mailto:services@ghamut.com}{services@ghamut.com}. \\
}

\usepackage{balance}       
\usepackage{graphics}      
\usepackage[T1]{fontenc}   
\usepackage{txfonts}
\usepackage{mathptmx}
\usepackage[pdflang={en-US},pdftex]{hyperref}
\usepackage{color}
\usepackage{booktabs}
\usepackage{textcomp}
\usepackage{xcolor}

\usepackage{microtype}        
\usepackage{ccicons}          

\usepackage{todonotes}

\def\plaintitle{Machine-arranged Interactions Improve Institutional Belonging and Cohesion \\}

\def\emptyauthor{}
\def\plainkeywords{Authors' choice; of terms; separated; by
  semicolons; include commas, within terms only; this section is required.}

\makeatletter
\def\url@leostyle{%
  \@ifundefined{selectfont}{
    \def\UrlFont{\sf}
  }{
    \def\UrlFont{\small\bf\ttfamily}
  }}
\makeatother
\def\pprw{8.5in}
\def\pprh{11in}

\definecolor{linkColor}{RGB}{6,125,233}
\hypersetup{%
  pdftitle={\plaintitle},
  pdfauthor={\emptyauthor},
  pdfkeywords={\plainkeywords},
  pdfdisplaydoctitle=true, 
  bookmarksnumbered,
  pdfstartview={FitH},
  colorlinks,
  citecolor=black,
  filecolor=black,
  linkcolor=black,
  urlcolor=linkColor,
  breaklinks=true,
  hypertexnames=false
}

\begin{document}

\title{\plaintitle}

\numberofauthors{3}
\author{%
  \alignauthor{Mohammad M. Ghassemi \\
    \affaddr{Ghamut Corporation}\\
    \email{ghassemi@ghamut.com}}\\
  \alignauthor{\textcolor{red}{  }}\\
  \alignauthor{Tuka Alhanai\\
    \affaddr{Ghamut Corporation}\\
    \email{tuka@ghamut.com}}\\
}

\maketitle

\begin{abstract}
We investigated how participation in machine-arranged meetings were associated with feelings of institutional belonging and perceptions of demographic groups. We collected data from 535 individuals who participated in a program to meet new friends. Data consisted of surveys measuring demography, belonging, and perceptions of various demographic groups at the start and end of the program. Participants were partitioned into a control group who received zero introductions, and an intervention group who received multiple introductions. For each participant, we computed twelve features describing participation status, demography and the amount of program-facilitated exposure to others who were similar to them and different from them. We used a linear model to study the association of our features with the participants' final belonging and perceptions while controlling for their initial belonging and perceptions. We found that those who participated in the machine-arranged meetings had 4.5\% higher belonging, and 3.9\% more positive perception of others.
\end{abstract}






\section{Introduction}

Organizations that foster belonging are more effective and better-able to retain their members \cite{gallup2013employee}. But belonging is not the only factor that improves an organization's effectiveness; studies report that diverse organizations are more creative than their less diverse counterparts \cite{stahl2010unraveling}. However, the positive impacts of diversity are only experienced by organizations with strong cross-group cohesion (i.e. individuals that are different, but also get along) \cite{gonzalez2009cross,horwitz2007effects}. 

For universities, students belonging and cross-group cohesion are particularly important. Students who feel they belong tend to be happier, higher-performing, and more successful alumni \cite{gallup2016purdueindex}. Students that experience a cohesive educational environment tend to make better employees and more effective leaders after graduation \cite{wolfe2013estimating}. Universities are well-aware of these advantages and invest considerable resources into improving campus diversity, inclusion, and belonging \cite{benokraitis2019affirmative}. Indeed, in 2018 alone, universities in the United States allocated between \$7 to \$10 million to initiatives promoting belonging and cohesion, an estimated two-fold increase since 2008 \cite{ohio2018diversity, au2018diversity, mich2018diversity}.

%
%

Current approaches to community programming at educational institutions are, for the most part, top-down: institutional officials are designated, who then thoughtfully organize programs and activities to improve belonging and cohesion. But given the immense demographic heterogeneity of most student bodies, programming may inadvertently (and ironically) exclude one set of students in an attempt to bring other sets together. Examples include: diversity barbeques that exclude vegans, diversity hiking trips that exclude the physically challenged and diversity pub-crawls that exclude those that don't drink.

Universities are aware that not all needs can be met through centrally-organized events and thus typically provide funding to support student-lead initiatives; the hope being that the full spectrum of students needs will be met through self-organization. While studies show that student-led initiatives do improve belonging, they do not necessarily improve cross-group cohesion \cite{shaulskiy2016belonging}; indeed, student groups tend to bring-together individuals who share common interests, backgrounds, or values \cite{karimi2018mapping}. It is uncommon to see student groups whose primary function is to bring together individuals who identify as different, rather than the same.

Hence, the primary challenges of organizing events that improve belonging and cohesion for heterogeneous student bodies are: (1) as the number of participants in an activity increases, so too does the probability of excluding any given participant, and (2) when given the option, students will almost always choose to associate with those that are more similar to, rather than different from them \cite{yamagishi2016parochial}. 

One solution to these challenges is to centrally organize many smaller events which occur with greater frequency where students may meet new peers in a context that may improve their feelings of belonging and perceptions of others that differ from them. However, the efficacy of such events has not been previously investigated. Indeed, prior investigations in this domain suffer from one or more of the following limitations: (1) they study online social interactions which may not reflect off-line social realities \cite{chen2009make, ducheneaut2006alone, spertus2005evaluating, zhao2012cultivating}, (2) they are purely observational, analysing associations between pre-existing relationships and outcomes without a `control group' for comparison \cite{hristova2016measuring, karimi2018mapping, morrison2008negative, peters2010social, pittman2008university}, (3) they use incomplete data-sets, lacking direct information on participants' feelings of belonging or their perceptions of others \cite{bakshy2015exposure,hristova2016measuring,karimi2018mapping,tondello2015chi}, and (4) they are cross-sectional, focusing on characterization of social networks and their association with outcomes, at only a single point in time \cite{ demanet2012school, hristova2016measuring, karimi2018mapping, peters2010social, sacco2014social, walton2012mere}.

In this study we seek to resolve the limitations of the prior work by investigating how \textit{in-person} participation in \textit{programmatically-arranged} one-on-one meetings between new students at a university campus were associated with their \textit{self-reported} feelings of institutional belonging and perceptions of others, \textit{over time}.

\section{Methods}

\subsection{Data Collection Platform} 
This study was approved by the Institutional Review Board at the University of Virginia. Data for this study were collected through an instance of the Connect platform operating at the University of Virginia's Charlottesville campus from September of 2018 through May of 2019. Connect is a web platform that facilitates one-on-one introductions between members of organizations. Users of the platform register online and are subsequently connected to other members of their institution with whom they have no prior acquaintance. Introductions are made once a week over email. The introductory email contains (1) the names of the individuals being introduced, (2) a suggested topic of conversation and (3) a recommended venue for an in-person meeting. The platform uses a combination of user-reporting and intermittent GPS tracking to verify if users actually meet in-person. Users of the Connect platform have no direct control over who they are introduced to; they can not, for instance, ``swipe" to update their match. More information on the Connect platform may be found online\footnote{\url{https://connectmaven.com}}. A copy of the email used to introduce participants may be found in the Appendix \footnote{Online Appendix: \url{https://ln2.sync.com/dl/264f44f60/dj9486yv-xdxj2jr8-a9fadfdc-5wa3p5a9}}.

\subsection{Participant Recruitment and Grouping} 
First year students were invited to join the Connect platform at the start of the Fall (2018) and Spring (2019) semesters, while second year and transfer students were invited to join in the Spring (2019) semester. Upon sign up, students were provided the choice to participate in a research study with the objective of improving their university experience. 
All registrants were made aware of the risks associated with meeting new people, and affirmed that they understood those risks and remained interested in participation. Students that agreed to participate as research participants in the Fall were randomly assigned to either an \textit{intervention}, or a \textit{wait-list} group. Members of the intervention group were introduced to three other students\footnote{Participants may have had additional introductions if a meeting was canceled.} selected at random without replacement during the Fall, while members of the wait-list group received no introductions during the Fall. All participants that joined in the Spring of 2019 were also randomly assigned to the intervention or wait-list group. In addition, all subjects that joined in the Fall of 2018, in both the intervention and wait-list groups, were introduced to three other students selected at random without replacement in the Spring.


\subsection{Surveys Collected} 
Research participants were asked to complete three types of surveys, each of which we describe here briefly: 
\begin{enumerate}

\item  \textit{Sign-up}: Upon initial registration to the program participants were invited to report their background (e.g. gender, race), personality traits (e.g. extroversion, dependability), perception of various groups (e.g. warmth towards Homosexuals, Muslims, Liberals), feelings of belonging, and levels of prior interaction with people from demographic backgrounds unlike their own. 
\item \textit{Post-lunch}: For each meeting logged on the platform, participants were invited to report their feelings about the person they met, their feelings of belonging, and their perceptions of various groups (e.g. Homosexuals, Muslims, Liberals).
\item \textit{End-of-semester}: At the end of each semester (Fall and Spring), participants were invited to report their feelings of belonging, their perceptions of various groups, and their impressions of the racial climate at their university. They also completed the Symbolic Racism Scale \cite{henry2002symbolic}.
\end{enumerate}

Additional details on each survey, including the specific questions asked, may be found in the online Appendix.

\subsection{Participation Incentive Policy} 
In exchange for their participation, participants were provided 10 USD for each meeting they attended, and 10 USD for completing surveys at the end of each semester. 

\subsection{Survey Simplification}
Surveys contained two varieties of questions. The first variety identified participants' \textit{categorical} membership within groups of interest (e.g. Gender: Male, Female) while the second variety identified the \textit{extent} of participants' membership within groups of interest (e.g. extroversion: 0 - 9). Below we describe our approach to simply the representations of the categorical and extent questions. 

\textit{Extent questions} were simplified from their seven- and nine-point representations to three-point representations \{-1,0,1\}. This was accomplished by first shifting the range of question values to be centered at zero. For example, extent questions ranging from 1 to 7 were shifted to range from -3 to 3. Next, negative (shifted) values were represented as -1, positive values as +1, and all other values as 0.

\textit{Categorical questions} were simplified differently depending on the phenomenon the question intended to describe. Regarding \textit{Gender}:  participants that identified as transgender,  non-conforming or 'another' were consolidated into a single group. Regarding \textit{Parental education}: terminal degree holders (M.D., Ph.D., J.D) were considered one group, non-terminal degree holders (BSc, MSc, Associates) were considered another group and high-school diploma holders were considered yet another group. Regarding \textit{Political Outlook}: liberals and conservatives were represented as -1 and +1 respectively with all others were represented as 0. We note that not all categorical questions were simplified; \textit{Sexual Orientation}, \textit{Race/Ethnicity} and \textit{Religion} retained the groupings assigned by the original survey (see online Appendix).

\subsection{Feature Representation}
Following the survey simplification approach described in the last section, we generated a set of twelve features that succinctly describe: the extent of each participant's membership in the demographic majority (one feature), their participation status in the program (one feature) and the similarities/differences between the participant and those they met in the program (five similarity features and five difference features). The selected features are shown in Table \ref{table:features}; we also describe the features briefly below. 

The \textit{participation feature} measured if the participant was in the intervention group (i.e. connected by the program) or was on the wait-list. The \textit{demographic feature} measured the extent to which the participant belonged to the demographic majority at the University of Virginia\footnote{Note, this is not necessarily the same as the demographic majority of those that participated in the program.}. \textit{Meeting features} measured the similarities/differences between the participant, and those they encountered, across five dimensions: Outgroup familiarity, Background, Belonging, Personality, and Group Perceptions. Five meeting features quantified the total number of \textit{similarities} between the participant and those they met during the program; five other meeting features quantified the total number of \textit{differences} between the participant and those they met in the program. That is, for each question that composes a given feature (e.g. gender, sexual orientation, race/ethnicity, etc. for the `background' feature) a similarity point is accumulated for each thing a participant had in common with a person they met (see Equation \ref{equation:similarity}); conversely a difference point was accumulated for everything they lacked in common with a person they met (see Equation \ref{equation:difference}).
\begin{equation}
    \label{equation:similarity}
    \text{similarity}(x,y) = 
    \begin{cases}
        1,& \text{if } x == y\\
        0,& \text{otherwise}
    \end{cases}
\end{equation}

\begin{equation}
    \label{equation:difference}
    \text{difference}(x,y) = 
    \begin{cases}
        1,& \text{if } x \neq y\\
        0,& \text{otherwise}
    \end{cases}
\end{equation}

If, for instance, two participants were female they would accumulate 1 background similarity point, otherwise they would accumulate 1 background difference point. Participants who attended no meetings were assigned no points of meeting similarities or differences. Hence, the meeting features also implicitly capture the number of meetings an individual had, since points are accumulated for a participant for every partner they met. This accumulation of points corresponds to the degree of exposure to individuals with similar or different profiles. 

We note that, at first glance, having both similarity and difference measures may seem redundant. However because each participant may attend a different number of meetings knowing only one of the measures does not make apparent the extent of the exposure in the other direction. That is, knowing the amount of similarity a participant was exposed to does not allow us to reconstruct the difference without additional information on the number of meetings a participant had \footnote{Albeit, the meeting features will be perfectly correlated for participants that attended only a single meeting}.

\begin{table*}[t]
    \centering
    \normalsize
    \begin{tabular}{|lcp{11cm}|}
\hline
\textbf{Participation Features (n = 1)} & \textbf{Range} & \textbf{Description}  \\ \hline
Participant Group  & {[}0 , 1{]} & 1 if participant was ever connected by the platform, else 0  \\ 
& & \\ \hline
\textbf{Demographic Features (n = 1)}   & \multicolumn{1}{l}{\textbf{Range}} & \textbf{Description} \\ \hline
Majority membership                     & {[}-2 - 2{]}                       & -2 + 1 for each of the following traits held by the participant: White, Male, Christian, Heterosexual.   \\ \hline
\textbf{Meeting Features (n = 10*)}     & \textbf{Range}  & \textbf{Description}  \\ \hline
Outgroup familiarity                    & {[}0 - 5$\lambda{]}$                        & The sum of the cardinality of the intersection of the sets that describe for each meeting the pair's contact with people of ethnicities that differ from their own: at college, as neighbors, as close friends and as informal acquaintances. \\ 
& &    \\
Background                              & {[}0 - 9$\lambda{]}$                        & The sum of the cardinality of the intersection of the sets that describe for each meeting the pair's: Gender, Sexual Orientation, Race, Religion, Political Views, Maternal Education Paternal Education, Year of Education, and Socioeconomic Status. \\
 &  & \\
Belonging                               & {[}0 - 5$\lambda{]}$                        & The sum of the cardinality of the intersection of the sets that describe for each meeting the pair's feelings of: belonging, being ``like others”, alienation, being informed and being accepted.  \\
& &  \\
Personality                             & {[}0 - 10$\lambda{]}$ & The sum of the cardinality of the intersection of the sets that describe for each meeting the pair's perceptions of their personalities as: critical/quarrelsome, extroverted/enthusiastic, dependable/disciplined, anxious/easily upset, open to new experiences/complex, reserved/quiet, sympathetic/warm, disorganized/careless, calm/emotionally stable, conventional/uncreative.  \\
                                        &                                    &  \\
Perceptions                             &  {[}0 - 29$\lambda{]}$             & Cardinality of the intersection of the sets that describe the pair's perceptions towards people from various groups: Teenagers, Young adults, Middle-aged adults, Older adults, Males, Females, Transgender or Gender Non-conforming people, Bisexuals, Gays/Lesbians, Queers, Heterosexual/Straight people, American Indians or Alaska Natives, Black or African Americans, East Asians, European American or White, Latino/as, Native Hawaiian or Pacific Islanders, Middle Easterners, South Asians, Christians, Jews, Buddhists, Hindus, Sikhs, Atheists, Agnostics, Liberals, Conservatives, Political Moderates \\ \hline
    \end{tabular}
   \caption{ Features that represent individual participation in the program. Features describe the level of participation, the extent of the participant's membership in the demographic majority, and the similarities/differences between the participant and those they met in the program. *Note that only five of the ten meeting features are displayed; the five unlisted features were similar to those listed, but were computed using the cardinality of the \textit{difference} (as opposed to the intersection) of the sets that describe the participants. $\lambda$ represents the number of meetings attended by the participant; it controls the upper-bound of the meeting feature range.}
   \label{table:features}
\end{table*}

\subsection{Outcome Representation}
Two outcomes were measured at the end of the program: (1) belonging and (2) warmth towards various demographic groups (which we will refer to hereafter as cohesion). In both cases, outcomes were based on participant's responses to a set of questions in the end-of-semester surveys. If participants completed only one end-of-semester survey, that survey's responses were used. If participants completed an end-of-semester survey in both the Fall of 2018 and the Spring of 2019, their Spring 2019 responses were used. Below we describe each of the outcomes, and the specific procedure used to generate them from the survey data.

\textit{Belonging} was computed using the participant's responses to the following questions in the end-of-semester survey: ($Q_1$) \textit{``I feel alienated from UVa"}, ($Q_2$) \textit{``People at UVa are a lot like me"}, ($Q_3$) \textit{"I belong at UVa"}  and ($Q_4$) \textit{ People at UVa accept me}. All questions were on a 7-point scale (with 1 indicating strong disagreement and 7 indicating strong agreement). In Equation \ref{equation:belonging_outcome}, we show the precise method of outcome computation:

\begin{equation}
    \label{equation:belonging_outcome}
    \text{Belonging} =  \frac{100}{28}(8 - Q_1 + Q_2 + Q_3 +Q_4)
\end{equation}

We note that the $(8 - Q_1)$ term in the above equation accounts for the negative orientation of $Q1$, and the $\frac{100}{28}$ scales the value of the feature to range between 0 and 100 (like a percentage).

\textit{Cohesion} was computed using the participant's responses to 31 questions ($W_{1:31}$) in the end-of-semester survey which asked them to report their warmth towards various age, gender, sexual orientation, religious, and political groups (See online Appendix for the complete list of questions). All questions were on an 11-point scale (with 0 indicating very cold and 10 very warm). In Equation \ref{equation:warmth}, we show the precise method of outcome computation:

\begin{equation}
    \label{equation:warmth}
    \text{Warmth} =  \frac{1}{310}\sum_{n=1}^{31}(W_n)
\end{equation}

We note that the $\frac{1}{310}$  term in the above equation accounts serves to scale the value of the feature between 0 and 100 (like a percentage).

\begin{table*}[]
\centering
\normalsize
\begin{tabular}{|lcccccc|}
\hline
\multicolumn{1}{|l|}{}                                                    & \multicolumn{1}{c|}{\textbf{Waitlisted}} & \multicolumn{5}{c|}{\textbf{\begin{tabular}[c]{@{}c@{}}Connected\end{tabular}}}                                  \\
\multicolumn{1}{|l|}{n = 535}                                               & \multicolumn{1}{c|}{n = 118}                 & n = 184                      & n = 128                        & n = 52                         & n = 35                         & n = 18                           \\ \hline
\multicolumn{1}{|l|}{\textbf{Meetings Attended}}                      & \multicolumn{1}{c|}{0}                     & 0                          & 1                            & 2                           & 3                            & \textgreater{}3      \\ \hline

\textbf{Demographics \% (n)}                                                              &                                                              &                              &                              &                              &                               &                               \\ \hline
\multicolumn{1}{|l|}{Gender - Female (384)}           & \multicolumn{1}{c|}{73\%}              & 69\%                     & 73\%                     &67\%                     & 76\%                     & 89\%                      \\
\multicolumn{1}{|l|}{Gender - Male (149)}               & \multicolumn{1}{c|}{27\%}              & 31\%                     & 27\%                     & 33\%                     & 24\%                     & 11\%                      \\ 
\hline

\multicolumn{1}{|l|}{Orientation - Heterosexual (454)}          & \multicolumn{1}{c|}{85\%}              & 87\%                     & 84\%                     & 86\%                     & 77\%                     & 94\%                     \\
\multicolumn{1}{|l|}{Orientation - Bisexual (47)}                 & \multicolumn{1}{c|}{9\%}                & 7\%                       & 10\%                     & 10\%                     & 14\%                     & 6\%                       \\
\multicolumn{1}{|l|}{Orientation - Homosexual/Other (30)} & \multicolumn{1}{c|}{6\%}                & 6\%                       & 6\%                      & 4\%                        & 9\%                      & 0\%                 \\ \hline
\multicolumn{1}{|l|}{Race - White (265)}                                & \multicolumn{1}{c|}{53\%}              & 43\%                     & 57\%                    & 52\%                     & 46\%                     & 44\%                     \\
\multicolumn{1}{|l|}{Race -  East Asian (74)}                        & \multicolumn{1}{c|}{10\%}              & 15\%                     & 13\%                    & 14\%                     & 20\%                     & 22\%                     \\
\multicolumn{1}{|l|}{Race -  South Asian (73)}                      & \multicolumn{1}{c|}{14\%}              & 16\%                     & 13\%                    & 10\%                     & 9\%                       & 11\%                     \\
\multicolumn{1}{|l|}{Race -  African American (39)}         & \multicolumn{1}{c|}{6\%}                & 9\%                       & 5\%                      & 6\%                       &  17\%                     & 0\%                     \\
\multicolumn{1}{|l|}{Race -  Middle Eastern (16)	}               & \multicolumn{1}{c|}{3\%}                & 4\%                       & 1\%                      & 4\%                       &  0\%                       & 6\%                     \\
\multicolumn{1}{|l|}{Race -  Latin American (11)}               & \multicolumn{1}{c|}{1\%}                & 4\%                       & 0\%                      & 2\%                       &  3\%                        & 0\%                     \\
\multicolumn{1}{|l|}{Race -  Other (5))}                          & \multicolumn{1}{c|}{1\%}                & 1\%                       & 0\%                      & 0\%                       &  0\%                        & 6\%                     \\ 
\hline

\multicolumn{1}{|l|}{Religion - Christian (215)}                     & \multicolumn{1}{c|}{39\%}              & 38\%                     & 43\%                     & 43\%                     & 46\%                     & 44\%                     \\
\multicolumn{1}{|l|}{Religion - Agnostic (120)}                     & \multicolumn{1}{c|}{29\%}              & 19\%                     & 23\%                     & 22\%                     & 23\%                     & 17\%                     \\
\multicolumn{1}{|l|}{Religion - Atheist (69)}                        & \multicolumn{1}{c|}{10\%}              & 18\%                     & 15\%                     & 4\%                       & 6\%                       & 17\%                     \\
\multicolumn{1}{|l|}{Religion - Another (35)}                     & \multicolumn{1}{c|}{6\%}                & 7\%                       & 6\%                        & 7\%                       & 7\%                       & 11\%                     \\ 
\multicolumn{1}{|l|}{Religion - Hindu (27)}                         & \multicolumn{1}{c|}{4\%}                & 6\%                       & 4\%                       & 8\%                       & 6\%                       & 0\%                     \\
\multicolumn{1}{|l|}{Religion - Muslims (21)}                     & \multicolumn{1}{c|}{4\%}                & 6\%                       & 2\%                       & 6\%                       & 0\%                       & 0\%                     \\
\multicolumn{1}{|l|}{Religion - Jewish (21)}                        & \multicolumn{1}{c|}{5\%}                & 2\%                       & 3\%                       & 6\%                       & 6\%                       & 11\%                     \\
\multicolumn{1}{|l|}{Religion - Buddhist (11)}                    & \multicolumn{1}{c|}{2\%}                & 2\%                       & 1\%                        & 4\%                       & 6\%                       & 0\%                     \\
\multicolumn{1}{|l|}{Religion - Sikh (8)}                           & \multicolumn{1}{c|}{1\%}                & 2\%                       & 3\%                        & 0\%                       & 0\%                       & 0\%                     \\
\hline

\multicolumn{1}{|l|}{Politics - Extremely Liberal (20)}                             & \multicolumn{1}{c|}{5\%}                 & 3\%                       & 5\%                      & 2\%                       & 3\%                     & 6\%                     \\
\multicolumn{1}{|l|}{Politics - Very Liberal (125)}                                      & \multicolumn{1}{c|}{22\%}               & 23\%                     & 27\%                    & 19\%                     & 28\%                   & 17\%                     \\
\multicolumn{1}{|l|}{Politics - Moderate Liberal (162)}                              & \multicolumn{1}{c|}{28\%}               & 35\%                     & 27\%                    & 26\%                     & 26\%                   & 22\%                     \\
\multicolumn{1}{|l|}{Politics - Slightly Liberal (62)}                                 & \multicolumn{1}{c|}{10\%}               & 11\%                     & 12\%                    & 12\%                     & 14\%                   & 22\%                     \\
\multicolumn{1}{|l|}{Politics - Neither (65)}     & \multicolumn{1}{c|}{13\%}               & 13\%                     & 9\%                      & 12\%                     & 17\%                   & 17\%                     \\
\multicolumn{1}{|l|}{Politics - Slightly Conservative (48)}                       & \multicolumn{1}{c|}{14\%}               & 7\%                       & 6\%                      & 15\%                      & 3\%                    & 11\%                     \\
\multicolumn{1}{|l|}{Politics - Moderately Conservative (40)}                 & \multicolumn{1}{c|}{6\%}                 & 6\%                       & 12\%                    & 8\%                       & 6\%                     & 0\%                     \\
\multicolumn{1}{|l|}{Politics - Very Conservative (13)}                            & \multicolumn{1}{c|}{2\%}                 & 2\%                       & 2\%                      & 6\%                       & 3\%                     & 5\%                     \\
\multicolumn{1}{|l|}{Politics - Extremely Conservative (0)}                  & \multicolumn{1}{c|}{0\%}                 & 0\%                       & 0\%                       & 0\%                       & 0\%                     & 0\%                     \\ 
\hline
\multicolumn{1}{|l|}{Student Status - First Year (392)}                              & \multicolumn{1}{c|}{42\%}              & 72\%                     & 72\%                     & 82\%                     & 85\%                     & 94\%                     \\ 
\multicolumn{1}{|l|}{Student Status - Second Year (164)}                         & \multicolumn{1}{c|}{58\%}              & 28\%                     & 28\%                     & 18\%                     & 15\%                     & 6\%                     \\
\multicolumn{1}{|l|}{Student Status - Transfer (9)}                               & \multicolumn{1}{c|}{1\%}                & 3\%                       & 1\%                       & 4\%                       & 0\%                       & 0\%                     \\ 
 \hline
\textbf{Meeting Features $\mu \,[\sigma]$}                                                                      &                                                              &                              &                             &                                &                              &                           \\ \hline
\multicolumn{1}{|l|}{Background - Similar}                          & \multicolumn{1}{c|}{0 [0]}             & 0 [0]                & 4.6 [1.4]                & 9.5 [2.5]                & 13.3 [3.7]     & 21.4 [6.0]       \\
\multicolumn{1}{|l|}{Background - Different}                        & \multicolumn{1}{c|}{0 [0]}             & 0 [0]                & 4.4 [1.4]                & 8.0 [2.2]                & 12.6 [4.1]     & 17.6 [4.7]       \\

\multicolumn{1}{|l|}{Personality - Similar}                         & \multicolumn{1}{c|}{0 [0]}             & 0 [0]                & 5.4 [1.4]                 & 11.0 [2.6]              & 16.1 [3.9]       & 24.4 [6.8]     \\
\multicolumn{1}{|l|}{Personality - Different}                       & \multicolumn{1}{c|}{0 [0]}             & 0 [0]                & 3.6 [1.4]                 & 6.4 [2.5]               & 9.8 [2.7]        & 14.6 [5.7]     \\

\multicolumn{1}{|l|}{Belonging - Similar}                           & \multicolumn{1}{c|}{0 [0]}             & 0 [0]                & 2.4 [1.3]                 & 4.8 [2.3]               & 7.7 [3.4]      & 11.4 [3.9]      \\
\multicolumn{1}{|l|}{Belonging - Different}                         & \multicolumn{1}{c|}{0 [0]}             & 0 [0]                & 1.6 [1.3]                 & 3.0 [2.2]               & 3.9 [3.3]      & 5.9 [4.0]       \\

\multicolumn{1}{|l|}{Perceptions - Similar}                         & \multicolumn{1}{c|}{0 [0]}             & 0 [0]                & 16.5 [9.5]                & 31.2 [16.1]             & 50.3 [25.2]     & 76.8 [23.5]    \\
\multicolumn{1}{|l|}{Perceptions - Different}                       & \multicolumn{1}{c|}{0 [0]}             & 0 [0]                & 10.5 [9.5]                & 21.3 [16.5]             & 27.7 [24.2]     & 40.1 [34.1]    \\

\multicolumn{1}{|l|}{Outgroup Familiarity - Similar}                & \multicolumn{1}{c|}{0 [0]}             & 0 [0]                & 3.2 [1.5]                 & 6.8 [2.5]               & 9.7 [4.1]       & 15.8 [5.6]    \\
\multicolumn{1}{|l|}{Outgroup Familiarity - Different}              & \multicolumn{1}{c|}{0 [0]}             & 0 [0]                & 1.8 [1.4]                 & 2.9 [2.1]               & 4.7 [3.8]       & 5.9 [4.2]    \\

 \hline

\textbf{Confounders $\mu \,[\sigma_{\hat{x}}]$}                                                                      &                                                              &                              &                             &                                &                              &                           \\ \hline
\multicolumn{1}{|l|}{Initial Belonging}                                                                      & \multicolumn{1}{c|}{70 [1.5]}            & 70 [1.1]                & 73 [1.4]                 & 76 [2.0]                 & 79 [1.8]       & 82 [24.7]                \\
\multicolumn{1}{|l|}{Initial Cohesion}                                                                         & \multicolumn{1}{c|}{64 [1.3]}             & 62 [1.3]                 & 61 [1.7]                  & 59 [2.9]                  & 60 [3.2]        & 69 [21.0]                    \\
 \hline
\textbf{Outcomes $\mu \,[\sigma_{\hat{x}}]$}                                                                      &                                                              &                              &                             &                                &                              &                           \\ \hline
\multicolumn{1}{|l|}{Final Belonging}                                                                      & \multicolumn{1}{c|}{74 [1.3]}             & 74 [0.9]                 & 75 [1.25]               & 78 [2.0]          & 80 [1.7]            & 82 [24.7]                      \\
\multicolumn{1}{|l|}{Final Cohesion}                                                                         & \multicolumn{1}{c|}{65 [1.3]}             & 60 [1.7]                  & 63 [1.6]                & 62 [2.7]           & 60 [3.0]             & 66 [19.7]   \\
 \hline
\end{tabular}
\caption{Demographics, exposures and outcomes of participants with respect to their status in the program, and the number of meetings attended.}
\label{table:demographics}
\end{table*}

\subsection{ Modeling Approach and Confounding Features}
We selected a linear modeling approach to ease interpretation and assessment of the relationship between our selected features and outcomes of interest. For each outcome, we trained a separate linear model and reported coefficients, 95\% confidence intervals, and p-values. We also reported the fit of our model using the $R^2$ metric.

Our selected outcomes reflect the belonging and cohesion of our participants at the \textit{end} of the program. We anticipated that these outcomes would be related to the participant's \textit{initial} feelings of belonging and cohesion, especially for those participants who did not participate. To account for these confounding effects, we generated two additional features measuring the \textit{initial} belonging and cohesion of the participants at the start of the program; the initial cohesion was included when modeling final cohesion and initial belonging was included when modeling final belonging. The features were computed using the same approach described above, with a corresponding set of questions from the participant's sign-up surveys. 

We suspected that measures of similarity and difference exposure for a given meeting feature (e.g. Similar Background, Dissimilar Background) may be correlated, especially if a significant number of participants only attend a single meeting. To prevent co-linearity, we decided to include \textit{either} the similarity or the difference measure in a given model. Furthermore, we hypothesized that (1) exposure to similarities were more likely to be positively associated with a participant's final belonging than exposure to differences and (2) exposure to differences were more likely to be positively associated with a participant's final cohesion than exposure to similarities. Hence, similarity features were used when modeling participant final belonging and difference features were used when modeling participant final cohesion. This approach facilitated a more natural interpretation of the model coefficients, and prevented the introduction of co-linear terms.

\section{Results}

\subsection{Demographics, Exposure and Participation}
A total of 562 users of the Connect platform agreed to participate as research participants; 299 participants signed-up in the Fall and 263 signed-up in the Spring. Of the 562 participants, 233 attended at least one meeting through the program, 84 were wait-listed in the Fall but participated in the Spring, 118 were wait-listed in the Spring (i.e. never connected by the platform) and 184 were connected but never logged a meeting. In total, 417 meetings were logged on the platform: 128 participants attended one meeting, 52 attended two meetings, 35 attended three meetings, 11 attended four meetings, 6 attended five meetings, and 1 attended six meetings. 

In Table \ref{table:demographics}, we compare the demographics, exposures and outcomes of participants with respect to their status in the program, and the number of meetings they attended. The participants were predominantly heterosexual (85\%) female (72\%), liberal (69\%), white (50\%) and Christian (40\%). The table shows that, in general, participants had more in common (i.e. similarities) with their matches than otherwise (i.e. differences). Participants that attended exactly one meeting were exposed to others with backgrounds, personalities, initial feelings of belonging, group perceptions, and out-group familiarity that were more likely to be similar to them than different from them (1.05x, 1.63x, 1.50x, 1.57x, and 1.78x respectively); participants that attended three meetings were even more likely to be exposed to others more similar to them than different from them (1.06x, 1.64x, 1.97x, 1.82x and 2.06x respectively).

For the most part, Table \ref{table:demographics} implies that the wait-listed participants, those which attended zero lunches, and those which attended one or more lunches were demographically similar. However, along some demographic axes, there were notable differences between the groups. The proportion of first-year participants in the intervention group, for instance, was much larger than the the proportion of first-year participants in the wait-list group \footnote{This was due to our preference towards first-year students as participants.}. Participants with higher cohesion were no more or less likely to attend additional meetings, nor was their final cohesion associated with the total number of meetings they reported attending. 

Table \ref{table:demographics} provides several interesting insights into factors that might be associated with final belonging: (1) participants with stronger initial feelings of belonging also attended more meetings, (2) improvements in belonging were more pronounced for those with lower initial feelings of belonging than for those with higher initial feelings and (3) final feelings of belonging were associated with initial feelings of belonging. 

\subsection{Confounders of Participation}
 In order to fairly assess the associations between program participation and our outcomes we must first understand and account for potential confounders of participation in the program. Hence, we performed a uni-variate Pearson's correlation between all factors from the sign-up survey, and the total number of meetings participants reported attending. After correcting for multiple hypothesis testing (Bonferroni  \cite{bland1995multiple}), three factors related to initial participant belonging exhibited a statistically significant Pearson's correlation with the number of lunches attended: initial feelings of belonging ($r = 0.24,  p < 0.01$), initial perceptions that others at UVa were ``a lot like them" ($r = 0.20, p < 0.01$), and initial feelings of being ``accepted" by others at UVa ($r = 0.15,  p < 0.01$). That is, the more strongly a participant felt they belonged initially, the more likely it was that they would attend multiple meetings with others. Importantly, all three of these factors are accounted for in the initial belonging feature, which was included as a confounder in our statistical models.


\begin{figure}[b!]
\centering
\includegraphics[width=0.99\columnwidth]{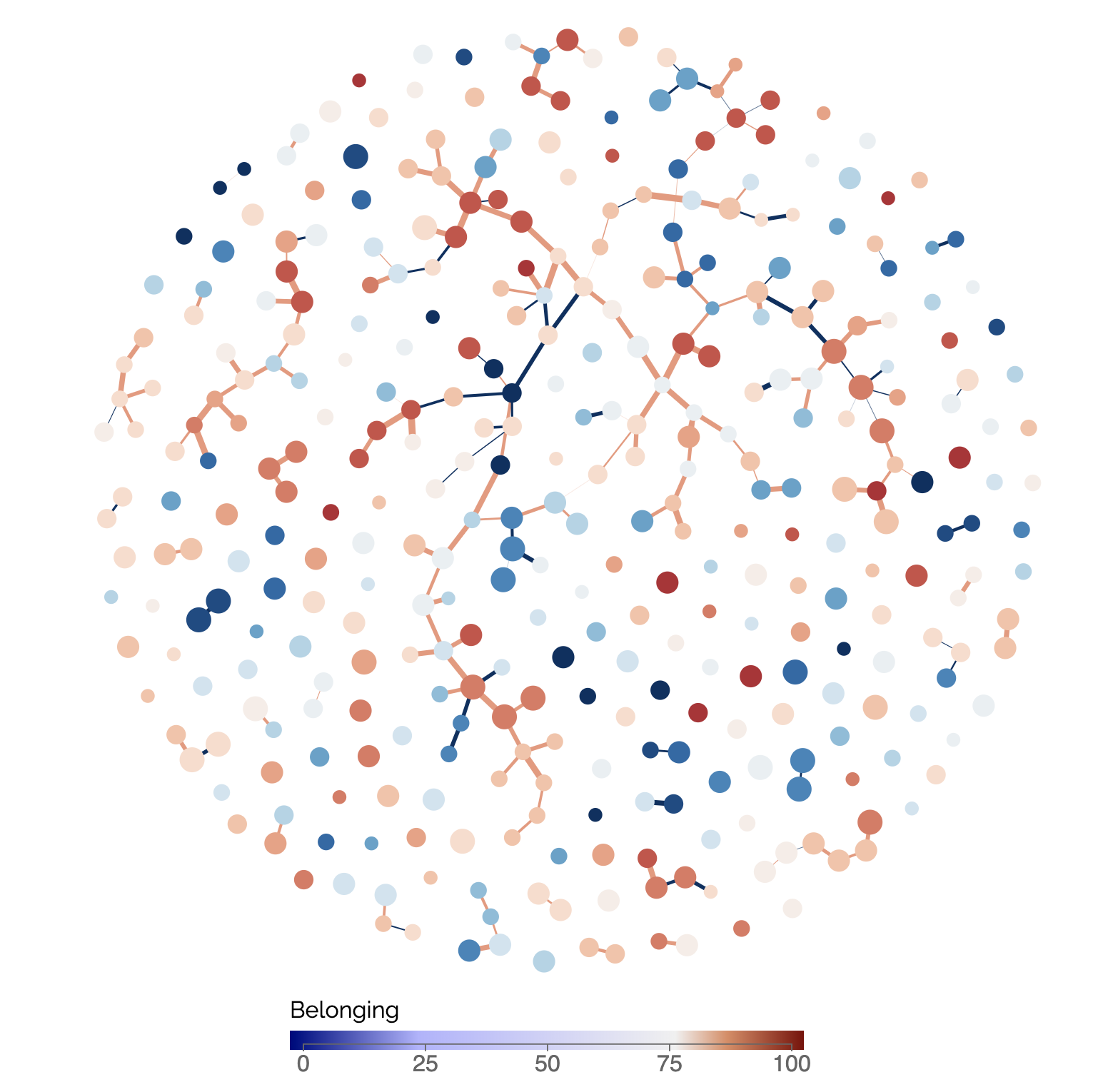}
\caption{Network plot of participants in the program as a function of their final belonging. Plot shows participants who logged a meeting or were on the wait-list (excluding those in the intervention group who never logged a meeting). \textbf{Node size} reflects extent in majority status. \textbf{Node color} reflects final belonging; redder nodes had \textit{higher} final belonging, bluer nodes had \textit{lower} final belonging. \textbf{Edge thickness} reflects the \textit{absolute value} of the cumulative difference of similarity and difference features for the pair of participants. \textbf{Edge color} indicates the \textit{polarity} of the difference between cumulative similarity and difference; a red edges indicates both participants were more \textit{similar} than different, and blue indicates that participants were more \textit{different} than similar. }
\label{fig:network-belonging}
\end{figure}

\subsection{Network Visualization}
\begin{figure}[t!]
\centering
\includegraphics[width=0.98\columnwidth]{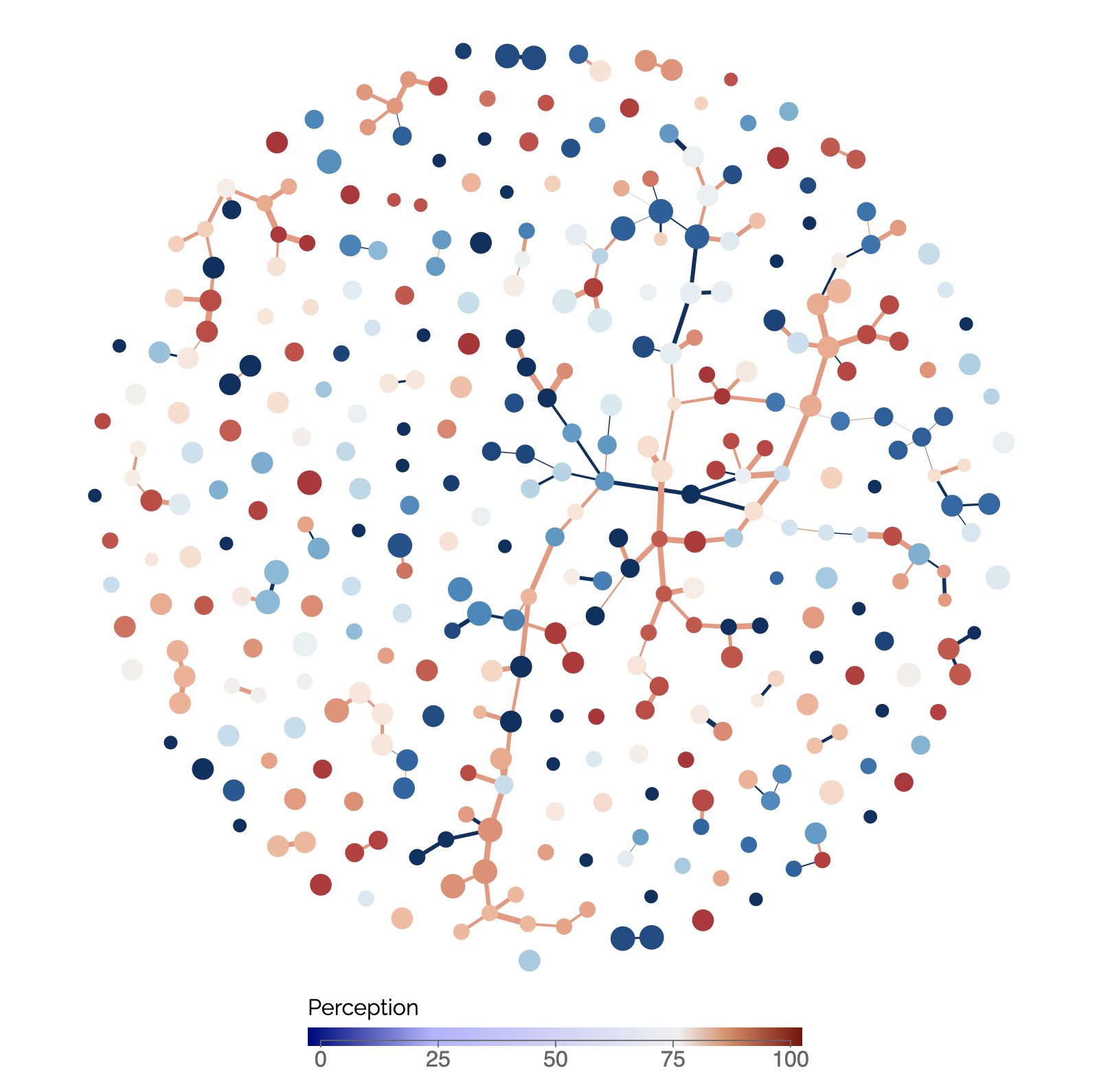}
\caption{Network plot of participants in the program as a function of their final cohesion towards other. Plot shows participants who logged a meeting or were on the wait-list (excluding those in the intervention group who never logged a meeting). \textbf{Node size} reflects extent in majority status. \textbf{Node color} indicates change in perception; Red nodes indicate \textit{increased} cohesion in their perception of the other group, Blue nodes indicates \textit{decreased} cohesion in their perception of the other group. \textbf{Edge color} indicates the \textit{polarity} of the difference between cumulative similarity and difference; a red edges indicates both participants were more \textit{similar} than different, and blue indicates that participants were more \textit{different} than similar.}
\label{fig:network-perception}
\end{figure}

In Figures \ref{fig:network-belonging} and \ref{fig:network-perception} we provide network plots that visualize the relationship between participation, meeting exposures, and our outcomes of interest (belonging and cohesion, respectively). At a glance, Figure \ref{fig:network-belonging} implies that wait-listed participants (isolated nodes) tended to have a lower final belonging scores (bluer) than those who participated. The Figure also implies that participants who attended more meetings (nodes with multiple edges) also generally tended to report higher values of final belonging (nodes were redder). Lastly, the figure implies that participants who attended more meetings also tended to have more in common with their matches than not (nodes had redder connecting edges).

Figure \ref{fig:network-perception} implies that wait-listed participants (isolated nodes) also tended to have a lower final cohesion (bluer nodes), albeit the relationship is less clear than the belonging outcome. Additionally, the figure implies that participants who attended more meetings (nodes with multiple edges) also tended have more in common with their matches than not (nodes had redder connecting edges).  Interestingly, we notice that the final cohesion of a node (color), was associated with the color of it's neighbors (i.e. blue nodes are more likely to be connected to blue, and red to red); this may imply that some participants were transmitting their feelings of cohesion to others.

\subsection{Statistical Associations}
Prior to statistical analysis, we excluded 50 participants who were missing sign-up or end-of-semester survey data required to generate our features and outcomes. Of the 485 remaining participants, 109 were wait-listed, 166 were connected but attended zero meetings, 115 attended one meeting 47 attended two meetings, 31 attended 3 meetings, 10 attended 4 meetings, 6 attended 5 meetings, 1 attended 6 meetings. The demographic, exposure, outcome and participation characteristics of the participants did not differ significantly from what was observed in Table \ref{table:demographics}. 

In Table \ref{table:belonging_model} we present a model that describes the association between our selected features and the belonging outcome ($R^2$ = 0.98). As expected, we found that initial feelings of belonging were strongly associated with the participants final feelings or belonging ($p < 0.01$). Indeed, a 1 unit increase in initial belonging was associated with a 0.99 unit increase in final belonging. After controlling for the effects of initial belonging, three other features exhibited a statistically significant positive association with final belonging: program participation ($Coeff:4.5\, p < 0.01$), exposure to others with similar backgrounds ($Coeff:0.52 \, p < 0.05$) and exposure to others with similar initial feelings of belonging ($Coeff: -0.91\, p < 0.03$). Given our results, we may claim with 99\% confidence that program participation was positively associated with final belonging. On average, those who participated had a final belonging 4.5\% higher than those on the wait-list. Furthermore, we may claim with 95\% confidence that for each element of demographic similarity a participant experienced on a meeting (e.g. meeting someone of the same religion), there was a 0.52\% increase in final reported belonging. Note that these effects are cumulative in nature; hence, for each person a participant met sharing all nine of the demographic traits measured in this study, we anticipate a 4.68\% absolute improvement in feelings of belonging. For each element of belonging similarity a participant experienced on a meeting (e.g. an alienated person meeting another alienated person) there was 0.91\% absolute decrease in final belonging (potential reasons for this association are provided in the discussion section).

\begin{table}[]
\begin{tabular}{lccc|}
\cline{2-4}
\multicolumn{1}{l|}{}                                             & \multicolumn{3}{c|}{\textbf{Final Belonging}}                                                             \\ \hline
\multicolumn{1}{|l|}{\textbf{Feature}}                            & Coeff.                  & 95\% C.I.                                    & p \textless{}=        \\ \hline
\multicolumn{1}{|l}{\textit{Participant Features:}}               &                              &                                              &                             \\ \hline
\multicolumn{1}{|l|}{{\color[HTML]{9A0000} Program Participation}}            & {\color[HTML]{9A0000} 4.5}   & {\color[HTML]{9A0000} {[} 2.27 , 6.76 {]}}   & {\color[HTML]{9A0000} 0.01} \\
\multicolumn{1}{|l|}{Majority Membership}                         & -0.48                        & {[} -1.44 , 0.49 {]}                         & 0.33                        \\ \hline
\multicolumn{1}{|l}{\textit{Confounders:}}                        &                              &                                              &                             \\ \hline
\multicolumn{1}{|l|}{{\color[HTML]{9A0000} Initial Belonging}}    & {\color[HTML]{9A0000} 0.99}  & {\color[HTML]{9A0000} {[} 0.96 ,1.01 {]}}    & {\color[HTML]{9A0000} 0.01} \\ \hline
\multicolumn{1}{|l}{\textit{Meeting Features:}}                   & \textit{}                    &                                              &                             \\ \hline
\multicolumn{1}{|l|}{{\color[HTML]{9A0000} Similar Background}}   & {\color[HTML]{9A0000} 0.52}  & {\color[HTML]{9A0000} {[} 0.01 , 1.05 {]}}   & {\color[HTML]{9A0000} 0.05} \\
\multicolumn{1}{|l|}{Similar Personality}                         & {\color[HTML]{333333} -0.29} & {\color[HTML]{333333} {[} -0.81 , 0.22 {]}}  & {\color[HTML]{333333} 0.28} \\
\multicolumn{1}{|l|}{{\color[HTML]{9A0000} Similar Belonging}}    & {\color[HTML]{9A0000} -0.91} & {\color[HTML]{9A0000} {[} -1.76 , -0.07 {]}} & {\color[HTML]{9A0000} 0.03} \\
\multicolumn{1}{|l|}{{\color[HTML]{333333} Similar  Perceptions}} & {\color[HTML]{333333} 0.06}  & {\color[HTML]{333333} {[} -0.05 , 0.16 {]}}  & {\color[HTML]{333333} 0.28} \\
\multicolumn{1}{|l|}{Similar Familiarity}                         & -0.09                        & {[}-0.63 , 0.45 {]}                          & 0.74                        \\ \hline
\end{tabular}
\caption{Coefficients, confidence intervals and statistical significance of an OLS Regression modeling the association between our selected features, and the belonging outcome. $R^2 = 0.98$.}
\label{table:belonging_model}
\end{table}

In Table \ref{table:warmth_model} we present a model which describes the association between our features and the cohesion outcome. Like belonging, we found that initial feelings of cohesion were strongly associated with the participants final feelings or cohesion ($p < 0.01$). Indeed, a 1 unit increase in initial cohesion was associated with a 0.91 unit increase in final cohesion. After controlling for the effects of initial cohesion, three other features exhibited a statistically significant association with the final cohesion: program participation ($Coeff:3.94\, p < 0.05$), exposure to others with different group perceptions ($Coeff:0.29 \, p < 0.01$) and exposure to others with different initial feelings of belonging ($Coeff: 1.29\, p < 0.05$). Given our results, we may claim with 95\% confidence that program participation was positively associated with final cohesion and that those who participated had 3.9\% higher cohesion than those who did not. Furthermore, we may claim with 99\% confidence that for each additional difference in group perceptions a participant experienced during a meeting (e.g. meeting someone who felt differently about Atheists than they did), there was a 0.29\% absolute increase in final reported cohesion. Note that these effects were cumulative in nature; hence, for each person a participant met with differences in all 29 perceptive traits measured in this study, we anticipate a 8.41\% absolute improvement in final cohesion. For each element of belonging difference a participant experienced on a meeting (e.g. an alienated person meeting someone who did not feel alienated) there was 1.29\% absolute increase in final cohesion (potential reasons for this association are provided in the discussion section).

\begin{table}[t!]
\centering
\begin{tabular}{lccc|}
\cline{2-4}
\multicolumn{1}{l|}{}                                               & \multicolumn{3}{c|}{\textbf{Final Cohesion}}                                                               \\ \hline
\multicolumn{1}{|l|}{\textbf{Feature}}                              & Coeff.                  & 95\% C.I.                                   & p \textless{}=        \\ \hline
\multicolumn{1}{|l}{\textit{Participant Features:}}                 &                              &                                             &                             \\ \hline
\multicolumn{1}{|l|}{{\color[HTML]{9A0000} Program Participation}}              & {\color[HTML]{9A0000} 3.94}  & {\color[HTML]{9A0000} {[} 0.01 , 7.87 {]}}  & {\color[HTML]{9A0000} 0.05} \\
\multicolumn{1}{|l|}{Majority Membership}                           & -0.26                        & {[} -1.96 , 1.44 {]}                        & 0.76                        \\ \hline
\multicolumn{1}{|l}{\textit{Confounders:}}                          &                              &                                             &                             \\ \hline
\multicolumn{1}{|l|}{{\color[HTML]{9A0000} Initial Cohesion}}         & {\color[HTML]{9A0000} 0.91}  & {\color[HTML]{9A0000} {[} 0.86 , 0.96 {]}}  & {\color[HTML]{9A0000} 0.01} \\ \hline
\multicolumn{1}{|l}{\textit{Meeting Features:}}                     & \textit{}                    &                                             &                             \\ \hline
\multicolumn{1}{|l|}{Different Background}                          & -0.35                        & {[} -1.16 , 0.46 {]}                        & 0.39                        \\
\multicolumn{1}{|l|}{Different Personality}                         & {\color[HTML]{333333} -0.32} & {\color[HTML]{333333} {[} -1.24 ,0.60 {]}}  & {\color[HTML]{333333} 0.49} \\
\multicolumn{1}{|l|}{{\color[HTML]{9A0000} Different Belonging}}    & {\color[HTML]{9A0000} 1.29}  & {\color[HTML]{9A0000} {[} 0.01 ,  2.60 {]}} & {\color[HTML]{9A0000} 0.05} \\
\multicolumn{1}{|l|}{{\color[HTML]{9A0000} Different  Perceptions}} & {\color[HTML]{9A0000} 0.29}  & {\color[HTML]{9A0000} {[} 0.11 , 0.47 {]}}  & {\color[HTML]{9A0000} 0.01} \\
\multicolumn{1}{|l|}{Different Familiarity}                         & -0.49                        & {[} -1.47 , 0.50 {]}                        & 0.33                        \\ \hline
\end{tabular}
\caption{Coefficients, confidence intervals and statistical significance of an OLS Regression modeling the association between our selected features, and the cohesion outcome. $R^2 = 0.92$.}
\label{table:warmth_model}
\end{table}

\section{Discussion, Limitations and Future Work}
Our results may be summarized as follows: program participation improved institutional belonging and cohesion; thoughtful exposure to others with similar demographic backgrounds, but different group perceptions, improved these outcomes all the more. With this in mind, we devote what remains of this paper to a discussion our our work, it's limitations and future directions.

The analyses presented in Tables \ref{table:belonging_model} and \ref{table:warmth_model} indicated that exposure to others with similar feelings of initial belonging was negatively associated with final belonging. This result prompted our further investigation into the relationship between the belonging exposure feature, and our outcomes. Interestingly, we found that when participants with below average belonging were exposed to others with below average belonging, their final belonging tended to decrease. However, when participants with above average belonging were introduced to others with above average belonging, there was no change in final belonging. Hence, the belonging similarity and difference features captured the negative effects of a low-belonging person meeting another low-belonging person.

Our modeling paradigm assumed that the effects of majority membership was independent from the effects of the other features. It is possible, however, that members of the demographic minority may react differently to the same meeting exposures than members of the demographic majority (e.g. people in the minority may react differently to meeting people like them relative to their peers in the majority). To explore this, we performed an additional analysis that included interaction terms between the majority extent feature and each of our meeting exposure features. None of the interactions were associated with the outcomes.

We decided not to include the number of meetings attended by the participants as a feature in the analysis for two reasons: (1) the meeting exposure features implicitly reflected (and correlated with) the total number of meetings attended by the participant and (2) we suspected that participants may not log all of their meetings through the platform. Regarding this latter point, we note that students self reported attending significantly more lunches than they logged on the platform \footnote{888 lunches: 263 reported meeting one person, 123 reported meeting two people,  57 reported meeting 3 people and 52 reported meeting 4 people}.

An interesting future direction for this work would be to quantify changes in belonging and cohesion after each meeting attended by the participants, rather than at the end of the program. Furthermore, it would be valuable to understand how the characteristics of a given meeting impact the probability of a participant attending a subsequent meeting (could people be 'scared off' by a bad meeting?). We did invite participants who logged meetings on the platform to report their feelings about the person they met in a post-lunch survey. However, only 33 individuals (18\% of those that logged at least one meeting) completed a post-lunch survey; due to the limited responses, information from the post-lunch surveys were excluded from our analysis.

To better understand how well our results will generalize to the greater university ecosystem, we intent to perform additional analyses over multiple years and across multiple institutions. This research objective will inevitably require partnership with our colleagues in the wider research community. To facilitate interest in this domain, and reproducibility of our results, we plan to release a deidentified version of the collected data, following publication.

We conclude this work by acknowledging that there were many things happening in student's lives that were not observed in our data including countless interactions with others that we are not privy to beyond what we prescribed to the participants via the Connect platform. Given this reality, it is remarkable and heartening that participation in the program had a measurable positive impact. 

\section{Conclusion}
We found that in-person participation in programmatically-arranged one-on-one meetings between new students at a university campus was associated with improvements in their self-reported feelings of institutional belonging, and perceptions of others. Thoughtful exposure to others with similar backgrounds, but different group perceptions, improved these outcomes all the more. Our findings highlight the potential of technology to make meaningful social impact and motivates additional research efforts into the utilization of technology to improve diversity and belonging in organizations. 

\section{Acknowledgements}
\noindent This work was generously supported by the Diversity \& Inclusion Grant from the University of Virginia and the Small Business Innovation Research (SBIR) Grant from National Science Foundation Award \#1845752 entitled ``SBIR Phase I: A Community-Building Platform That Uses Machine Learning To Facilitate In-Person Interactions." The Authors would like to acknowledge Dr. Timothy Wilson and Dr. Sophie Trawalter for their collaboration on this work.

\balance{}

\bibliographystyle{SIGCHI-Reference-Format}
\bibliography{main}

\end{document}